\pdfoutput=1
\documentclass[cernpreprint, texlive=2023, english, texmf, orcidlogo]{atlasdoc}

\usepackage{atlaspackage}
\usepackage{atlasbiblatex}
 
\usepackage{atlasphysics}

\graphicspath{{logos/}{figures/}}
 
\usepackage{ANA-HIGG-2021-11-PAPER-defs}

\AtlasTitle{Study of High-Transverse-Momentum Higgs Boson Production in Association with a Vector Boson in the  $qqbb$ Final State with the ATLAS Detector}

\AtlasAbstract{
This Letter presents the first study of Higgs boson production in association with a vector boson ($V = W$ or $Z$) in the fully hadronic $qqbb$ final state using data recorded by the ATLAS detector at the LHC in proton--proton collisions
at $\sqrt{s}=13$~TeV and corresponding to an integrated luminosity of 137~fb$^{-1}$. The vector bosons and Higgs bosons are each reconstructed as large-radius jets and tagged using jet substructure techniques. Dedicated tagging algorithms exploiting $b$-tagging properties are used to identify jets consistent with Higgs bosons decaying into $b\bar{b}$.
Dominant backgrounds from multijet production are determined directly from the data, and a likelihood fit to the jet mass distribution of Higgs boson candidates is used to extract the number of signal events.
The $VH$ production cross section is measured inclusively and differentially in several
ranges of Higgs boson transverse momentum: 250--450, 450--650, and greater than 650~GeV.
The inclusive signal yield relative to the standard model expectation is observed to be $\mu = 1.4 ^{+1.0}_{-0.9}$ and the corresponding cross section is $3.1 \pm 1.3\, (\text{stat.})\: ^{+1.8}_{-1.4}\, (\text{syst.}$)~pb.
}
 
\author{The ATLAS Collaboration}

\PreprintIdNumber{CERN-EP-2023-261}

\AtlasJournalRef{\PRL   132 (2024) 131802}
\AtlasDOI{10.1103/PhysRevLett.132.131802}

\hypersetup{pdftitle={ATLAS document},pdfauthor={The ATLAS Collaboration}}
 
\begin{document}
 
\maketitle

In the Standard Model (SM) of particle physics, the Brout--Englert--Higgs mechanism~\cite{Englert:1964aa,Higgs:1964aa,Higgs:1964ab,Guralnik:1964aa} spontaneously breaks electroweak symmetry. As a result, a physical Higgs boson emerges and both the $W$ and $Z$ gauge bosons and the fermions acquire mass.
Since the discovery of the Higgs boson by the ATLAS and CMS experiments at the LHC~\cite{HIGG-2012-27,CMS-HIG-12-028}, precision measurements of its properties have been a priority for both experiments.
The Higgs boson decay into bottom-quark pairs ($H\rightarrow b\bar{b}$) has the largest branching fraction ($\sim$58\%)~\cite{LHCHXSWG} and has only recently been observed by exploiting associated production with a vector boson ($VH$, $V=W, Z$)~\cite{HIGG-2018-04,CMS-HIG-18-016}.  
In these analyses, the associated $W$ and $Z$ bosons were required to decay leptonically to provide both a means to trigger on these events and an effective suppression of the large background from QCD production of $b$-quarks.
Theoretical work has raised interest in testing for new physics contributions to associated $VH$ production at high momentum by exploiting $H\rightarrow b\bar{b}$ decays in a highly boosted event topology~\cite{Brehmer:2019gmn,Biekotter:2014gup,Mimasu:2015nqa} because of its sensitivity to higher-order effective operators. On the experimental side, ATLAS and CMS have successfully developed and calibrated novel jet-substructure techniques to identify high-momentum $V$ or $H$ bosons that decay hadronically and are reconstructed as a single large-radius (\enquote{large-$R$}) jet. Dedicated tagging algorithms exploiting $b$-tagging properties are used to identify jets consistent with Higgs bosons decaying into $b\bar{b}$.
 
In this paper, these experimental techniques are applied in a first study of associated $VH$ production at high Higgs boson transverse momentum (\pTH) in the two large-$R$ jets topology.
This approach has the potential to probe higher \pTH values than in the best current measurement in the semileptonic final
state~\cite{HIGG-2018-52} because the $W/Z$ hadronic branching fractions are larger.
The $VH$ signal is extracted using a likelihood fit to the invariant mass distribution of Higgs candidate large-$R$ jets in events also containing a $W$- or $Z$-candidate jet. The dominant multijet background is estimated directly from the data, while smaller backgrounds ($t\bar{t}$, $V$+jets, and $VV$ production) are modeled using  Monte Carlo (MC) simulation. The normalization of the peaking $Z$+jets background (mostly from $Z \to b\bar{b}$ decays passing as Higgs candidates) is a free parameter in the final fit. 
The selected events are separated into several \pTH\ ranges and the $VH$ production cross section is measured both inclusively and as a function of \pTH.
 
The analysis presented here uses the proton--proton collision data collected by the ATLAS detector~\cite{PERF-2007-01,ATLAS-TDR-19,PIX-2018-001} from 2015 to 2018, corresponding to an integrated luminosity of 137~\fb\ at $\rts = 13$~TeV.
The ATLAS detector is a multipurpose particle detector with cylindrical geometry.
It consists of an inner tracking detector surrounded by a superconducting solenoid, sampling electromagnetic and hadronic calorimeters, and a muon spectrometer with three toroidal superconducting magnets, providing a near 4$\pi$ coverage in solid angle.\footnote{ATLAS uses
a right-handed coordinate system with its origin at the nominal interaction point in the center of
the detector and the $z$-axis along the beam pipe. The $x$-axis points to the center of the LHC ring,
and the $y$-axis points upwards. Cylindrical coordinates ($r, \phi$) are used in the transverse plane, $\phi$ being
the azimuthal angle around the $z$-axis.
The pseudorapidity is defined in terms of the polar angle $\theta$ as $\eta = -\ln\tan(\theta/2)$.
Angular distance is measured in units of $\Delta R \equiv \sqrt{(\Delta \eta)^2+(\Delta \phi)^2}$.}
A two-level trigger system~\cite{TRIG-2016-01} selects events for storage.
An extensive software suite~\cite{ATL-SOFT-PUB-2021-001} is used in data simulation, in the reconstruction and
analysis of real and simulated data, in detector operations, and in the trigger and data acquisition systems of the experiment.
Collision events satisfy a number of requirements ensuring that the ATLAS detector was operating well while the data were recorded.
 
Simulated events were produced for a variety of processes.  All generated events were passed through a full simulation of the ATLAS detector response~\cite{SOFT-2010-01}  using \GEANT~\cite{Agostinelli:2002hh}. The effects of multiple $pp$ interactions in the same
or neighboring bunch crossings (\pileup) were included by overlaying events generated with \PYTHIA[8]~\cite{Sjostrand:2014zea}.
Events were weighted such that the distribution of the average number of interactions per bunch crossing matches that observed in data.

Higgs boson events were generated at next-to-leading-order (NLO) accuracy via gluon--gluon fusion (ggF), vector-boson fusion (VBF), and in association with a vector boson ($VH$) using \POWHEGBOX[v2]~\cite{Nason:2009ai,Alioli:2010xd,Nason:2004rx,Frixione:2007vw} and the \PDFforLHC[15nlo] parton distribution function (PDF) set~\cite{Butterworth:2015oua}. The loop-induced $gg\to ZH$ process was generated separately at leading order (LO) with \POWHEGBOX.
In these cases, the events were interfaced with \PYTHIA[8]~\cite{Sjostrand:2014zea} utilizing the \AZNLO tune~\cite{STDM-2012-23} to incorporate parton shower and nonperturbative effects.
The production of a Higgs boson in association with two top quarks ($\ttbar H$) was modeled similarly, using \POWHEGBOX[v2] with the \NNPDF[3.0nlo] PDF set~\cite{Ball:2014uwa} and the A14 tune~\cite{ATL-PHYS-PUB-2014-021}.
Other Higgs boson production mechanisms such as $tH$ and QCD-induced $b\bar{b}H$ are not considered since their contribution is negligible.
The simulated processes are normalized using the best available cross-section calculations~\cite{LHCHXSWG}. 
In particular, the $pp \to VH$ cross section is calculated to next-to-next-to-leading order (NNLO) in QCD with NLO electroweak (EW) corrections, and the $gg \to ZH$ cross section is calculated at NLO and next-to-leading-logarithm (NLL) accuracy~\cite{Ciccolini:2003jy,Brein:2003wg,Brein:2011vx,Altenkamp:2012sx,Denner:2014cla,Brein:2012ne,Harlander:2014wda}.
The $q\bar{q}/ qg\to ZH$ simulated events are normalized using the $pp \to ZH$ cross section (which combines the $q\bar{q}$, $qg$, and $gg$ contributions) after subtracting the $gg \to ZH$ cross section.
Differential NLO EW corrections computed with \textsc{HAWK}~\cite{Denner:2014cla} are applied to the $q\bar{q}/ qg \to VH$ process as a function of vector-boson transverse momentum, $\pt^V$.

Simulated multijet events are only used to optimize the event selection since the multijet background is estimated directly from the data, as described later. They were generated using \PYTHIA[8] with LO matrix elements for dijet production. 
The \NNPDF[2.3lo] PDF set~\cite{Ball:2012cx} and A14 tune were used in all steps of the event generation.
Samples of $V$+jets events were generated with \SHERPA[2.2.8]~\cite{Gleisberg:2008ta} and the \NNPDF[3.0nlo] PDF set at NLO accuracy for one additional parton and LO accuracy for up to four additional partons.
All generated $V$+jets samples have NLO QCD and LO EW accuracy. Additional NLO EW corrections are applied to the simulated events as weights.
Diboson events ($VV = WW, WZ, ZZ$) were generated with \SHERPA[2.2.8] and the \NNPDF[3.0nlo] PDF set at NLO accuracy for up to one additional parton and LO accuracy for up to three additional partons.
The simulated \ttbar\ events were generated at NLO using \POWHEGBOX[v2] with $\hdamp=1.5 m_\text{top}$~\cite{ATL-PHYS-PUB-2016-020}. 
Weights are also applied to match the \ttbar\ yields measured in data in the same kinematic phase space~\cite{HIGG-2021-08}.
Single-top production in the $Wt$ channel was modeled at NLO using \POWHEGBOX[v2] in the five-flavor scheme~\cite{Frixione:2008yi} and with the diagram removal scheme~\cite{ATL-PHYS-PUB-2016-020} to treat interference with \ttbar\ production. The \NNPDF[3.0nlo] PDF set was used for all top-quark samples.
Except for \SHERPA samples, the decay of $b$- and $c$-hadrons was performed using \EVTGEN[1.6.0]~\cite{Lange:2001uf}.

Events are required to pass single large-$R$ jet triggers~\cite{TRIG-2016-01}, with mass and transverse momentum exceeding specific thresholds. The event selection described below is designed to ensure full trigger efficiency. Events are also required to have a primary vertex with at least two associated tracks.
The primary vertex is selected as the vertex with the largest \(\Sigma \pT^2\), where the sum is over all of its associated tracks
with \(\pT > \SI{0.5}{\GeV}\)~\cite{ATL-PHYS-PUB-2015-026}.
In addition, noncollision backgrounds originating from calorimeter noise, beam halo interactions or cosmic rays are suppressed by
rejecting events that contain any \akt calorimeter jet with $R=0.4$ failing to satisfy a set of quality criteria~\cite{ATLAS-CONF-2015-029}.
 
Large-$R$ jets are used to identify both the $V$ boson and the Higgs boson. They are reconstructed with the \akt algorithm~\cite{Cacciari:2008gp, Fastjet} with a radius parameter of $R=1.0$, utilizing locally weighted topological cell clusters~\cite{PERF-2014-07} for their constituents. At this stage, these jets
are referred to as ungroomed jets. Jets are then trimmed~\cite{Krohn:2009th} to remove energy deposits from \pileup and the underlying event as well as soft and wide-angle radiation. The jet mass is reconstructed by combining the calorimeter and tracking measurements~\cite{JETM-2018-02}, where tracks are required to be associated with the primary vertex.
 
To aid in \Hbb tagging of large-$R$ jets, a separate collection of jets is built from tracks with the \akt\ algorithm using a \pt-dependent radius~\cite{ATL-PHYS-PUB-2017-010}. Variable-radius track-jets are then matched to \largeR calorimeter jets via ghost association~\cite{Cacciari:2008gn} and used as proxies for $b$-quarks associated with the Higgs candidate. All large-$R$ jets must satisfy $\pT > 200$~GeV and $|\eta|<2.0$ to ensure they are contained within the tracking detector.
Events with one or more isolated charged leptons (electrons or muons) are rejected. Electrons are identified  by matching tracks to energy clusters in the electromagnetic calorimeter.  They must have $\pt> 7$~GeV and  $|\eta|<2.5$, and satisfy the \enquote{loose} identification criterion defined in Ref.~\cite{EGAM-2018-01}. Muon identification relies on matching tracks in the inner detector to muon spectrometer tracks or track segments. Muons must have $\pt > 7$~GeV and $|\eta| < 2.5$, and satisfy the \enquote{loose} selection criterion~\cite{MUON-2018-03}.

The identification of $W$ and $Z$ bosons relies on several properties of the large-$R$ jets. The \DTwoBetaOne variable exploits the two-prong structure of the $W/Z\rightarrow qq$ decays, absent in typical QCD jets~\cite{Larkoski:2014gra,Larkoski:2015kga}.
The number of tracks ($N_\text{trk}$) linked to the ungroomed large-$R$ jets by ghost association is significantly higher for gluon-induced jets in background events than for quark-induced jets in signal events, due to the distinct energy scales involved and the different color factors for gluons and quarks. Requirements on the jet mass, \DTwoBetaOne, and $N_\text{trk}$ are optimized as a function of jet \pt\ such that a signal efficiency of 50\%~\cite{ATL-PHYS-PUB-2020-017} is achieved in each jet-\pt\ category, along with a multijet background rejection factor of 70--200.
The $W$-boson tagging is calibrated with semileptonic \ttbar\ data events and then extrapolated to the identification of $Z$ bosons~\cite{ATL-PHYS-PUB-2020-017}. For the high-\pt category, an additional uncertainty is applied to the $Z$-boson tagging, based on differences between $W$-boson to $Z$-boson tagging efficiency extrapolations obtained with \Sherpa and \Herwig.
 
Higgs candidates are identified by an \Hbb tagger~\cite{ATL-PHYS-PUB-2020-019} based on a neural-network algorithm that uses track and vertex information from the variable-radius track-jets to discriminate between large-$R$ jets from Higgs boson decays into $b\bar{b}$ and jets from gluons, light quarks, or top quarks.
A fixed 60\% signal efficiency working point (WP) is used, where 60\% refers to the average efficiency for selecting simulated Higgs bosons with $\pt > 250$~GeV~\cite{ATL-PHYS-PUB-2020-019}.
The jet mass resolution for Higgs candidates is improved by applying a \enquote{muon-in-jet} correction to account for the energy carried by muons from semileptonic $b$-hadron decays~\cite{PERF-2017-04}. The corrected jet mass of the Higgs boson candidate (\mJH) is used to extract the signal.

Events are required to contain at least two selected large-$R$ jets. 
The leading (highest \pT ) large-$R$ jet must have $\pT > 450$~GeV and mass above 50~GeV, to ensure full trigger efficiency. The second leading large-$R$ jet must have a mass above 40~GeV. At least one of those two jets must pass the \Hbb tagger requirements.
If both jets satisfy those requirements, the one with a larger mass is selected as the Higgs candidate. The other jet must then satisfy the $W/Z$ tagging requirements.
Furthermore, the Higgs candidate's transverse momentum (\pTJH) must exceed 250~GeV, thereby defining the signal region (SR), and three exclusive ranges are chosen: $\pTJH \in [250,450)$~GeV, $\pTJH \in [450,650)$~GeV, and $\pTJH \geq 650$~GeV.
In the SR, the $VH$ process dominates (85\%), although other Higgs boson production processes contribute: \ttH (8\%), ggF (6\%), and  VBF (1.4\%). The dominant background contribution comes from multijet production (90\%), followed by $t\bar{t}$ (5\%), $V$+jets (3.6\%), and diboson (0.7\%) production.
A data-driven method is used to estimate the background contribution from multijet production as well as that from $V$+jets production in which the $W/Z$ boson is correctly tagged and the additional jet is a QCD jet misidentified as the Higgs candidate (both processes are referred to as \enquote{multijet}). Other background processes are modeled by simulation, including $V$+jets production in which the $W/Z$ boson is selected as the Higgs candidate and a QCD jet passes the $V$ tagging requirements.

The multijet background estimate is extracted from a control region (CR) where events pass all SR requirements except the Higgs tagging requirement, since the \Hbb tagger does not rely on the jet mass as a discriminating variable. To account for differences in distribution shape and normalization between the SR and CR, a transfer factor (TF), which depends on both jet \pt and jet mass ($m$), is used.  The number of multijet events in the SR is thus derived from the number of multijet events in the CR ($N_\text{multijet}^\text{CR}$) as
\begin{equation*}
N^\text{SR}_\text{multijet}(\pt, m) = \text{TF} (p_\text{T}, \rho) \times N_\text{multijet}^\text{CR}(\pt, m),
\end{equation*}
where $\text{TF} (p_\text{T}, \rho) = \sum_{k,l}{\alpha_{kl} \: \rho^k\pt^l}$, $\rho=\log(m^2/\pt^2)$, and $\alpha_{kl}$ are the polynomial coefficients for the $k$-th order in $\rho$ and $l$-th order in \pt. The $\alpha_{kl}$ coefficients are determined from a simultaneous fit to the
data in the SR and CR across the whole jet mass range.  To determine the order of the polynomial needed to fit the data, a Fisher $F$-test is performed. Based on its results, a first-order polynomial in both $\rho$ and \pT is found to be sufficient to parameterize the transfer factor in the SR.

Three validation regions were defined to verify that the multijet background estimated using the TF method describes the observed background well. Events in these regions contain a $V$-jet passing the looser $W/Z$ tagging 80\% efficiency WP~\cite{ATL-PHYS-PUB-2020-017} but failing either the $N_\text{trk}$, \DTwoBetaOne, or mass requirement corresponding to the nominal 50\% efficiency WP. The composition of events from background processes in these regions, as tested with simulated samples, is similar to that in the SR. The background modeling is found to be in good agreement with data within the statistical uncertainty in each of the validation regions.

A second multijet background estimation method is utilized to test the nominal TF method. 
A boosted decision tree (BDT), trained on an alternative set of data events that fail both the $V$ and Higgs boson tagging, is used to perform a kinematic reweighting to match the kinematic distributions in this alternative region to those in the SR. A description of this reweighting procedure is available in Ref.~\cite{HDBS-2018-11}.
The two multijet background estimates agree within statistical uncertainties.

Systematic uncertainties arise from several different sources: the data-driven background estimate, the experimental reconstruction, and the theoretical prediction for the signal.
Their impact is summarized in the Appendix.
The uncertainty affecting the shape of the estimated multijet background is evaluated as the difference between the distributions from the BDT and TF methods. It is found to be up to 10\% for $\pTJH \in [250, 650)$~GeV and up to 20\% for \pTJH\ above 650~GeV.

Uncertainties in the modeling of the subdominant $VV$ and \ttbar\ backgrounds include normalization uncertainties of 80\%~\cite{HIGG-2018-52} and 12\%~\cite{TOPQ-2016-09}, respectively, as well as changes in the shape of the \mJH\ distribution when using different renormalization and factorization scales or alternative event generators.
The effect of scale uncertainties on the \mJH\ distribution's shape is also included for the $Z$+jets background.
Uncertainties in the modeling of the $Wt$ background have negligible impact.
 
Experimental uncertainties related to large-$R$ jets, $W/Z$ tagging, and \Hbb tagging affect the predicted event yields for the signal and the $VV$, $V$+jets, and \ttbar\ backgrounds. Uncertainties in the scale and resolution of the \largeR jet energy and mass measurements affect the shape of the \mJH\ distribution and are evaluated following Ref.~\cite{JETM-2018-02}.
 
Uncertainties in the $W/Z$ tagging are determined from studies of \ttbar\ events in data~\cite{JETM-2018-03,ATL-PHYS-PUB-2020-017}.
The \Hbb tagging efficiency is measured in a sample enriched in $Z \to b\bar{b}$ decays~\cite{ATL-PHYS-PUB-2021-035}. The resulting scale factors are applied to the simulation and vary between 0.86 and 1.80 across the \pt~bins, with uncertainties in the range 30\%--60\%.
These scale factors are further constrained in the fit to data by the presence of the $W/Z$ resonances in the jet mass distribution.
 
The impact of uncertainties in the PDF sets, initial- and final-state radiation, and multiparton interactions on the signal
acceptance is included. Uncertainties related to the PDF sets are derived by applying the methodology
outlined by the PDF4LHC group~\cite{Butterworth:2015oua} and considering four additional PDF sets (CT14, MMHT2014,
NNPDF3.0, and ATLAS-epWZ12), resulting in 3\% uncertainties in the signal acceptance.
An uncertainty of 0.83\% is applied to the integrated luminosity~\cite{ATLAS:2022hro}.
 
\begin{figure}[t]
\includegraphics[width=\textwidth]{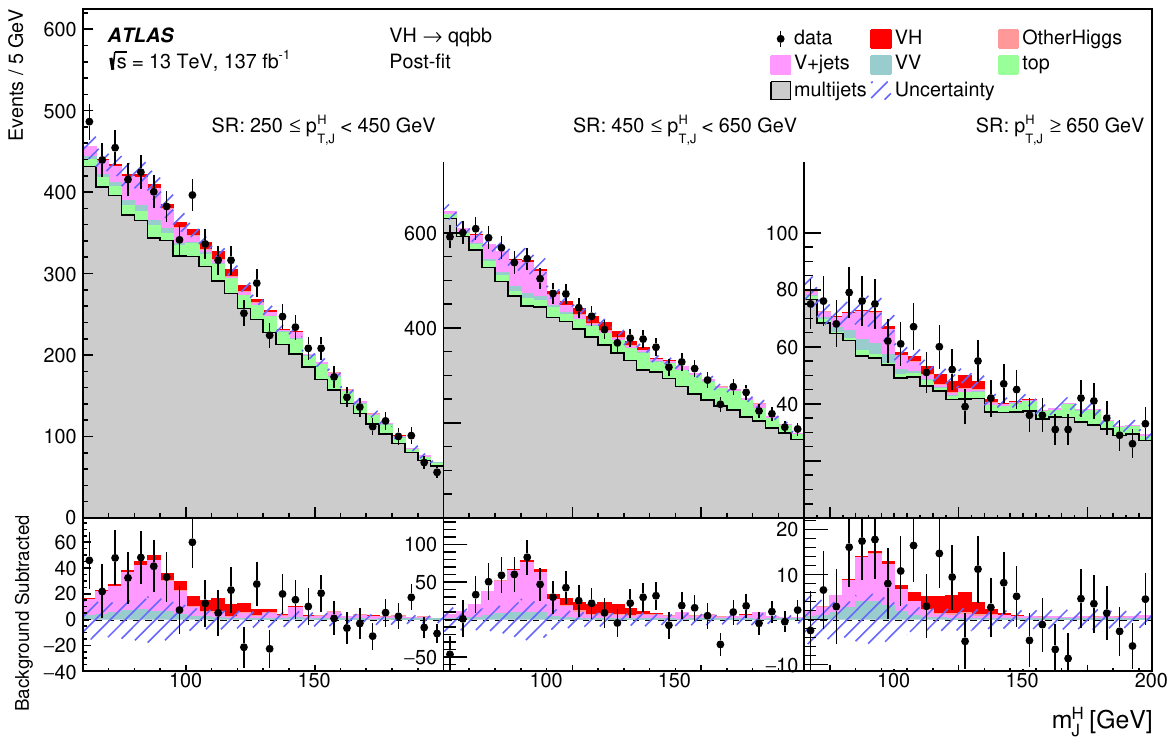}
\caption{Higgs candidate jet mass distributions in the signal region for $\pTJH \in [250,450)$~GeV (left), $\pTJH \in [450,650)$~GeV (middle), and $\pTJH \geq 650$~GeV (right) obtained after the inclusive fit with a single $Z$+jets normalization factor and a single signal strength. 
The bottom panels show the distributions after subtracting the multijet and top-quark backgrounds. The hatched bands show to the total uncertainty in the background estimate.
}
\label{fig:MHfit}
\end{figure}

The signal yield and the $Z$+jets normalization are extracted from a simultaneous binned maximum-likelihood fit to the \mJH distributions in the SR (shown in Figure~\ref{fig:MHfit}) and CR in the range 60--200~\GeV.
Confidence intervals are based on the profile-likelihood-ratio test statistic~\cite{Kendall-Stuart,Cowan:2010js}. Systematic uncertainties are implemented in the fit as nuisance parameters constrained by Gaussian or log-normal likelihood terms, and the Higgs boson mass is assumed to be $m_H = 125.09 \pm 0.24~\GeV$. The overall normalization factor for the peaking component of the $Z$+jets background (mostly from $Z \to b\bar{b}$ decays) from the fit is $1.4 ^{+0.8}_{-0.6}$.
 
The best-fit value of the signal-strength parameter, defined as the ratio of the observed signal yield to that expected in the SM, is $\mu = 1.4 ^{+1.0}_{-0.9}$ for the inclusive fit, corresponding to an observed (expected) significance of $1.7\,\sigma$ ($1.2\,\sigma$) with respect to the null signal hypothesis. An inclusive cross section of 2.24~pb~\cite{cepeda2019higgs} is used to normalize the expected signal.
The statistical uncertainty in $\mu$ is 0.6, whereas the systematic uncertainty is $^{+0.8}_{-0.6}$. The latter is dominated by the shape of the estimated multijet background and the \Hbb tagging scale factors.
The corresponding inclusive cross section is $3.1 \pm 1.3\, (\text{stat.})\: ^{+1.8}_{-1.4}\, (\text{syst.}$)~pb.
 
\begin{table}[th]
\centering
\caption{Signal strengths ($\mu$) and cross sections ($\sigma$) in exclusive kinematic regions. The expected cross sections are based on the inclusive cross section from Ref.~\cite{cepeda2019higgs} and the acceptance values derived from the signal simulation. The upper limits on the cross section provided in parentheses are quoted at the 95\% confidence level.}
\label{tab:cross}
\renewcommand{\arraystretch}{1.3}
\begin{tabular}{lcrlc}
\hline \hline
Kinematic region & Observed $\mu$ & \multicolumn{2}{c}{Observed $\sigma$ [fb]} & Expected $\sigma$ [fb] \\
\hline
$250 \le \pTH < 450$~GeV, $|y_H|<2$ & $0.8^{+2.2}_{-1.9}$ & $47^{+125}_{-109}$ & ($<$363) & 57.0 \\
$450 \le \pTH < 650$~GeV, $|y_H|<2$ & $0.4^{+1.7}_{-1.5}$ & $2^{+10}_{-9}$~~ & ($<$~~24) & ~~5.9 \\
$\pTH \ge 650$~GeV, $|y_H|<2$ & ~$5.3^{+11.3}_{-3.2}$ & $6^{+13}_{-4}$~~ & ($<$~~43) & ~~1.2 \\
\hline \hline
\end{tabular}
\end{table}

The signal strengths and cross sections are also extracted in three exclusive kinematic regions defined at generator level by $\pTH \in [250,450)$~GeV, $\pTH \in [450,650)$~GeV, and $\pTH \geq 650$~GeV, with $|y_H| < 2$ required in each region.
The results are obtained using \mJH\ templates extracted from the signal MC samples with the kinematic requirements applied for each region. Those results are presented in Table~\ref{tab:cross}.

In conclusion, a first study of associated $VH$ production is performed in the fully hadronic final state, reaching Higgs boson transverse momenta at the TeV scale.
A likelihood fit to the mass distribution of Higgs candidate large-$R$ jets in events also containing a $W$ or $Z$ candidate jet is used to extract the $VH$ signal both inclusively and as a function of transverse momentum. The significance of the $VH$ signal is estimated to be $1.7\, \sigma$ and
the inclusive cross section is determined to be $3.1 \pm 1.3\, (\text{stat.})\: ^{+1.8}_{-1.4}\, (\text{syst.}$)~pb, in agreement with the SM prediction.
While the current study is limited by large uncertainties, this channel will open a kinematic region with high sensitivity to new physics contributions when larger data samples are collected.

\FloatBarrier

\section*{Acknowledgments}

% The next lines are included from the .//acknowledgements/Acknowledgements.tex input file

We thank CERN for the very successful operation of the LHC, as well as the
support staff from our institutions without whom ATLAS could not be
operated efficiently.
 
We acknowledge the support of
ANPCyT, Argentina;
YerPhI, Armenia;
ARC, Australia;
BMWFW and FWF, Austria;
ANAS, Azerbaijan;
CNPq and FAPESP, Brazil;
NSERC, NRC and CFI, Canada;
CERN;
ANID, Chile;
CAS, MOST and NSFC, China;
Minciencias, Colombia;
MEYS CR, Czech Republic;
DNRF and DNSRC, Denmark;
IN2P3-CNRS and CEA-DRF/IRFU, France;
SRNSFG, Georgia;
BMBF, HGF and MPG, Germany;
GSRI, Greece;
RGC and Hong Kong SAR, China;
ISF and Benoziyo Center, Israel;
INFN, Italy;
MEXT and JSPS, Japan;
CNRST, Morocco;
NWO, Netherlands;
RCN, Norway;
MEiN, Poland;
FCT, Portugal;
MNE/IFA, Romania;
MESTD, Serbia;
MSSR, Slovakia;
ARRS and MIZ\v{S}, Slovenia;
DSI/NRF, South Africa;
MICINN, Spain;
SRC and Wallenberg Foundation, Sweden;
SERI, SNSF and Cantons of Bern and Geneva, Switzerland;
MOST, Taipei;
TENMAK, T\"urkiye;
STFC, United Kingdom;
DOE and NSF, United States of America.
In addition, individual groups and members have received support from
BCKDF, CANARIE, CRC and DRAC, Canada;
PRIMUS 21/SCI/017 and UNCE SCI/013, Czech Republic;
COST, ERC, ERDF, Horizon 2020, ICSC-NextGenerationEU and Marie Sk{\l}odowska-Curie Actions, European Union;
Investissements d'Avenir Labex, Investissements d'Avenir Idex and ANR, France;
DFG and AvH Foundation, Germany;
Herakleitos, Thales and Aristeia programmes co-financed by EU-ESF and the Greek NSRF, Greece;
BSF-NSF and MINERVA, Israel;
Norwegian Financial Mechanism 2014-2021, Norway;
NCN and NAWA, Poland;
La Caixa Banking Foundation, CERCA Programme Generalitat de Catalunya and PROMETEO and GenT Programmes Generalitat Valenciana, Spain;
G\"{o}ran Gustafssons Stiftelse, Sweden;
The Royal Society and Leverhulme Trust, United Kingdom.
 
The crucial computing support from all WLCG partners is acknowledged gratefully, in particular from CERN, the ATLAS Tier-1 facilities at TRIUMF/SFU (Canada), NDGF (Denmark, Norway, Sweden), CC-IN2P3 (France), KIT/GridKA (Germany), INFN-CNAF (Italy), NL-T1 (Netherlands), PIC (Spain), RAL (UK) and BNL (USA), the Tier-2 facilities worldwide and large non-WLCG resource providers. Major contributors of computing resources are listed in Ref.~\cite{ATL-SOFT-PUB-2023-001}.

% End of text imported from the .//acknowledgements/Acknowledgements.tex input file

\clearpage
\appendix
\part*{Appendix}

\begin{figure}[b]
\includegraphics[width=0.98\textwidth]{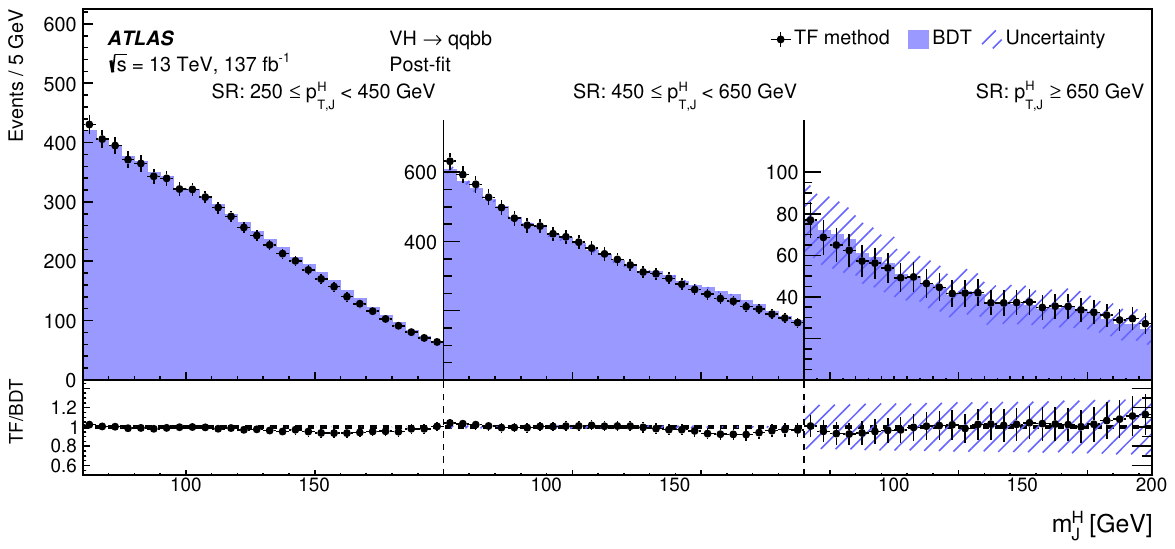}
 
\caption{Higgs candidate jet mass distributions for the multijet background in the signal region estimated with either the nominal transfer factor method (TF) or the boosted decision tree method (BDT). The error bars (hatched uncertainty band) represent(s) the total uncertainty in the TF (BDT) estimate including both the statistical and systematic uncertainties. The BDT uncertainties comprise a statistical component obtained using the method from Ref.~\cite{HDBS-2018-11} and the difference between the data and the background estimate in the validation regions. This difference is relatively large in the last \pt bin. It should be noted that there is a strong statistical correlation between the distributions from the two methods and that the BDT uncertainties are small in the first two \pt bins.}
\label{fig:BDTvsTF}
 
\end{figure}

The Monte Carlo simulation does not provide an accurate description of the QCD multijet background, especially in the high-\pT kinematic region, motivating the need for a dedicated estimate of the multijet contribution in the phase space selected for this study.
Therefore, this analysis exploits a fully data-driven estimation of the multijet background, using events in both the signal and control regions while testing it in validation regions.
Two different methods are compared for the data-driven estimation:
\begin{itemize}
\item The transfer factor (TF) method 
uses a jet \pT- and mass-dependent transfer function to predict the yields of events that pass the event selection from the events that fail the Higgs boson tagging. The multijet background estimation and the signal extraction are performed simultaneously.
 
\item The boosted decision tree (BDT) method 
extracts the background templates from the events failing the $V$ and Higgs boson tagging. A BDT is used to perform a kinematic reweighting, by predicting the event weights needed to bring the shapes of kinematic distributions in the control and signal regions into agreement. The following variables are used in the BDT training: Higgs candidate large-$R$ jet \pt, mass, pseudorapidity and azimuthal angle, as well as the number of associated tracks and track-jets.
 
\end{itemize}
The two methods give consistent results. After studying the fit to the sidebands and the uncertainties in the signal-plus-background Asimov fit~\cite{Cowan:2010js} to the signal region, the TF method was selected as the background estimation method, while the BDT method provides a robust and important cross-check of the results obtained and is used as an alternative method to determine shape systematic uncertainties. A comparison of the Higgs candidate jet mass distributions for the multijet background estimates from the two methods is shown in Figure~\ref{fig:BDTvsTF}.

\begin{figure}[]
\includegraphics[width=0.98\textwidth]{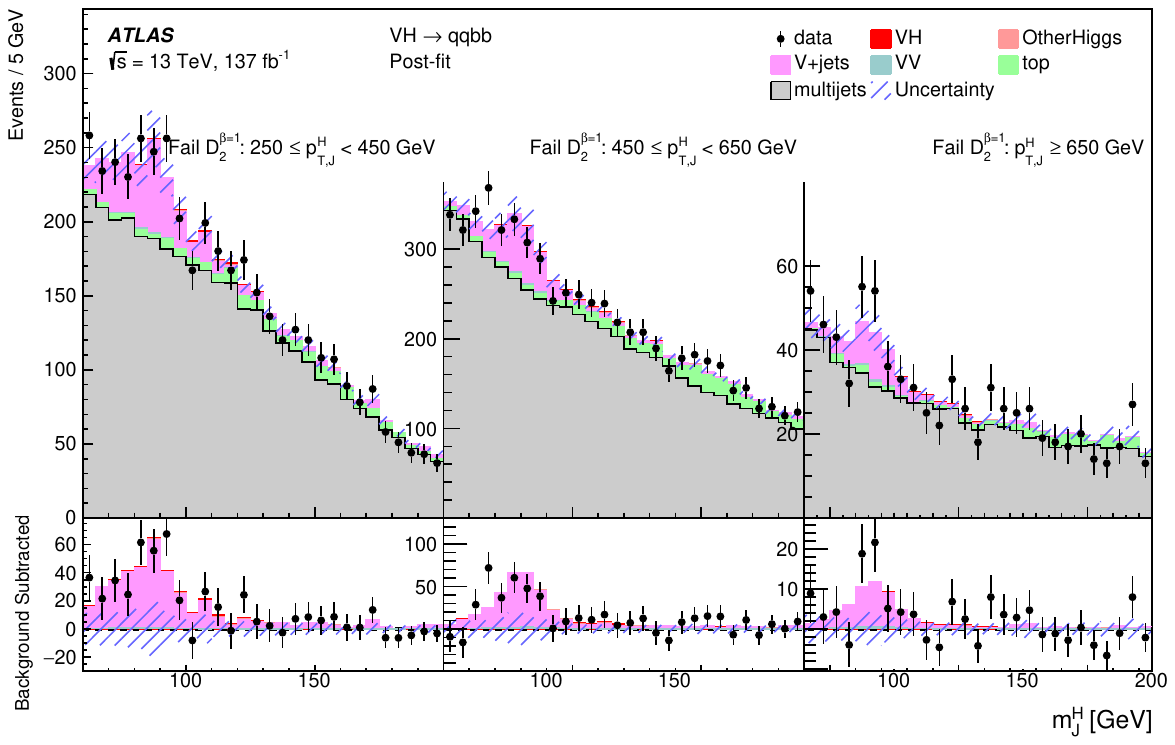}
 
\caption{Higgs candidate jet mass distributions after the inclusive fit in the validation region consisting of events passing the 80\% efficiency working point of the $V$-jet tagger but failing only the \DTwoBetaOne requirement of its tighter 50\% efficiency working point. No \Hbb tagging is applied. The bottom panels show the distributions after subtracting the multijet and top-quark backgrounds. The hatched bands show to the total uncertainty in the background estimate.}
\label{fig:VR_failD2}
 
\end{figure}

As another test of the background estimation, the statistical analysis is applied to the validation and control regions. Figure~\ref{fig:VR_failD2} presents the Higgs candidate jet mass distributions in one of the validation regions and demonstrates good agreement between the data and the background model.

The impact of the different sources of uncertainty on the signal strength is presented in Table~\ref{tab:uncertainties}.
 
\begin{table}[h]
\caption{Breakdown of the various sources of uncertainty in the inclusive signal strength $\mu$.
}
\centering
\renewcommand{\arraystretch}{1.4}
\begin{tabular}{ll}
\toprule
Uncertainty source   & $\delta\mu$ \\ \hline
\midrule
 
Signal modeling                    &  $^{+0.10}_{-0.02}$ \\
MC statistical uncertainty         &  $^{+0.13}_{-0.13}$ \\
Instrumental (pileup, luminosity)  &  $^{+0.012}_{-0.004}$ \\
Large-$R$ jet                      &  $^{+0.13}_{-0.14}$ \\
Top-quark modeling                 &  $^{+0.14}_{-0.15}$ \\
Other theory modeling              &  $^{+0.05}_{-0.03}$ \\
\Hbb tagging                       &  $^{+0.52}_{-0.23}$ \\
Multijet estimate (TF uncertainty) &  $^{+0.52}_{-0.41}$ \\
Multijet modeling (TF vs.\ BDT)     &  $^{+0.14}_{-0.18}$ \\ \hline
Total systematic uncertainty       &  $^{+0.80}_{-0.61}$ \\ \hline\hline
Signal statistical uncertainty     &  $^{+0.60}_{-0.60}$ \\
$Z$+jets normalization             &  $^{+0.42}_{-0.20}$ \\ \hline
Total statistical uncertainty      &  $^{+0.63}_{-0.63}$ \\ \hline\hline
Total uncertainty                  &  $^{+1.02}_{-0.88}$ \\
\bottomrule
 
\end{tabular}
\label{tab:uncertainties}
\end{table}

\clearpage
\printbibliography

\clearpage
% ATLAS Collaboration author list
% Reference date of HIGG-2021-11 is 2023-08-26
% Author list last updated on date 23-FEB-24
% Data extracted on 23-Feb-2024 for paper reference HIGG-2021-11
% at 9:47am
 
\begin{flushleft}
\hypersetup{urlcolor=black}
{\Large The ATLAS Collaboration}

\bigskip

\AtlasOrcid[0000-0002-6665-4934]{G.~Aad}$^\textrm{\scriptsize 102}$,
\AtlasOrcid[0000-0002-5888-2734]{B.~Abbott}$^\textrm{\scriptsize 120}$,
\AtlasOrcid[0000-0002-1002-1652]{K.~Abeling}$^\textrm{\scriptsize 55}$,
\AtlasOrcid[0000-0001-5763-2760]{N.J.~Abicht}$^\textrm{\scriptsize 49}$,
\AtlasOrcid[0000-0002-8496-9294]{S.H.~Abidi}$^\textrm{\scriptsize 29}$,
\AtlasOrcid[0000-0002-9987-2292]{A.~Aboulhorma}$^\textrm{\scriptsize 35e}$,
\AtlasOrcid[0000-0001-5329-6640]{H.~Abramowicz}$^\textrm{\scriptsize 151}$,
\AtlasOrcid[0000-0002-1599-2896]{H.~Abreu}$^\textrm{\scriptsize 150}$,
\AtlasOrcid[0000-0003-0403-3697]{Y.~Abulaiti}$^\textrm{\scriptsize 117}$,
\AtlasOrcid[0000-0002-8588-9157]{B.S.~Acharya}$^\textrm{\scriptsize 69a,69b,m}$
% ...
\bigskip
\\

% ...
$^{29}$Physics Department, Brookhaven National Laboratory, Upton NY; United States of America.\\
% ...
$^{35}$$^{(a)}$Facult\'e des Sciences Ain Chock, R\'eseau Universitaire de Physique des Hautes Energies - Universit\'e Hassan II, Casablanca;$^{(b)}$Facult\'{e} des Sciences, Universit\'{e} Ibn-Tofail, K\'{e}nitra;$^{(c)}$Facult\'e des Sciences Semlalia, Universit\'e Cadi Ayyad, LPHEA-Marrakech;$^{(d)}$LPMR, Facult\'e des Sciences, Universit\'e Mohamed Premier, Oujda;$^{(e)}$Facult\'e des sciences, Universit\'e Mohammed V, Rabat;$^{(f)}$Institute of Applied Physics, Mohammed VI Polytechnic University, Ben Guerir; Morocco.\\
% ...
$^{49}$Fakult\"{a}t Physik , Technische Universit{\"a}t Dortmund, Dortmund; Germany.\\
% ...
$^{55}$II. Physikalisches Institut, Georg-August-Universit\"{a}t G\"ottingen, G\"ottingen; Germany.\\
% ...
$^{69}$$^{(a)}$INFN Gruppo Collegato di Udine, Sezione di Trieste, Udine;$^{(b)}$ICTP, Trieste;$^{(c)}$Dipartimento Politecnico di Ingegneria e Architettura, Universit\`a di Udine, Udine; Italy.\\
% ...
$^{102}$CPPM, Aix-Marseille Universit\'e, CNRS/IN2P3, Marseille; France.\\
% ...
$^{117}$Department of Physics, New York University, New York NY; United States of America.\\
% ...
$^{120}$Homer L. Dodge Department of Physics and Astronomy, University of Oklahoma, Norman OK; United States of America.\\
% ...
$^{150}$Department of Physics, Technion, Israel Institute of Technology, Haifa; Israel.\\
$^{151}$Raymond and Beverly Sackler School of Physics and Astronomy, Tel Aviv University, Tel Aviv; Israel.\\
% ...
$^{a}$ Also Affiliated with an institute covered by a cooperation agreement with CERN.\\
$^{b}$ Also at An-Najah National University, Nablus; Palestine.\\
$^{c}$ Also at Borough of Manhattan Community College, City University of New York, New York NY; United States of America.\\
$^{d}$ Also at Center for High Energy Physics, Peking University; China.\\
$^{e}$ Also at Center for Interdisciplinary Research and Innovation (CIRI-AUTH), Thessaloniki; Greece.\\
$^{f}$ Also at Centro Studi e Ricerche Enrico Fermi; Italy.\\
% ...
$^{m}$ Also at Department of Physics, King's College London, London; United Kingdom.\\
% ...

\end{flushleft}

% Created with Glance <Atlas.Glance@cern.ch>

\end{document}